\documentclass[11pt]{article}
\usepackage{amssymb,latexsym,amsmath,amsbsy}
\usepackage[dvips]{graphicx}
\usepackage{cite}
\newtheorem{theo}{Theorem}

\newtheorem{lemm}[theo]{Lemma}
\newtheorem{coro}[theo]{Corollary}

\newtheorem{prop}[theo]{Proposition}
\def\nn{\nonumber}

\def\qdots{\mathinner{\mkern1mu\raise1pt\vbox{\kern7pt\hbox{.}}\mkern2mu
 \raise4pt\hbox{.}\mkern2mu\raise7pt\hbox{.}\mkern1mu}}

\def\h{\mathfrak{h}}
\def\so{\mathfrak{so}}

\def\osp{\mathfrak{osp}}

\begin{document}
\begin{center}
{\Large \bf
Representations of the orthosymplectic Lie superalgebra $\osp(1|4)$ and paraboson coherent states} \\[2mm]
{\bf R.~Chakrabarti\footnote{E-mail: ranabir@imsc.res.in; Permanent address: 
Department of Theoretical Physics, University of Madras, Guindy Campus, Chennai 600 025,
India}, }
{\bf N.I.~Stoilova}\footnote{E-mail: Neli.Stoilova@UGent.be; Permanent address:
Institute for Nuclear Research and Nuclear Energy, Boul.\ Tsarigradsko Chaussee 72,
1784 Sofia, Bulgaria} {\bf and J.\ Van der Jeugt}\footnote{E-mail:
Joris.VanderJeugt@UGent.be}\\[1mm]\
Department of Applied Mathematics and Computer Science,
Ghent University,\\
Krijgslaan 281-S9, B-9000 Gent, Belgium.
\end{center}
\centerline{\bf{Abstract}}
We introduce and obtain multimode paraboson coherent states. In appropriate 
subspaces these coherent states provide a decomposition of 
unity where the measure, when expressed using the cat-type states, is positive 
definite. Bicoherent states where the mutually commuting lowering 
operators are diagonalized are also obtained. Matrix elements in the coherent 
state basis are calculated. 
\setcounter{equation}{0}
\section{Introduction} \label{sec:Introduction}%

Soon after parastatistics has been introduced~\cite{Green}, it was discovered that it has a deep algebraic
structure. It turned out that any $n$ pairs of parafermion operators generate the simple Lie algebra
$\so(2n + 1)$\cite{Kamefuchi,Ryan}, and $n$ pairs of paraboson creation and annihilation operators 
$b_1^\pm, \ldots,b_n^\pm$
generate a Lie superalgebra~\cite{Omote}, isomorphic to one of the basic classical Lie superalgebras
 in the classification of Kac~\cite{Kac}, namely to the orthosymplectic Lie superalgebra
$\osp(1|2n) $~\cite{Ganchev}. Actually, the paraboson operators 
were introduced earlier by Wigner~\cite{Wigner} in a search of the most general commutation relations between
the position operator $\hat q$ and the momentum operator $\hat p$ of a one-dimensional oscillator, so that the
Heisenberg equations are compatible with  Hamilton's equations. The operators $\hat p,\; \hat q$ turned out to
generate $\osp(1|2)$ and Wigner was the first who found a class of (infinite-dimensional) representations
of a Lie superalgebra~\cite{Palev1}. Later on the results of Wigner gave rise to more general quantum
systems (see~\cite{Ohnuki} for references in this respect and for a general introduction to parastatistics), and
in particular to Wigner quantum systems introduced by Palev~\cite{Palev1,Palev2}.
 
However only quite recently, the paraboson Fock spaces for the Lie superalgebra $\osp(1|2n)$ were constructed~\cite{paraboson}. 
These are lowest weight representations $V(p)$ characterized by a positive parameter $p$, called the order 
of the statistics. In~\cite{paraboson}, an explicit basis for the representation spaces $V(p)$ is introduced
with the matrix elements of these representations. Because of the computational difficulties in the construction 
of the paraboson Fock spaces the paraboson coherent states (eigenstates of paraboson operators) 
were constructed only for one pair of paraboson 
operators~\cite{Sharma}. In the present paper we use the results of~\cite{paraboson} for the $n=2$ case in order to 
obtain ``coherent state'' representations of two pairs of paraboson operators $b_1^\pm$, $b_2^\pm$. Coherent states play vital roles~\cite{KS85, P86, Ali, Vourdas} 
in many contexts such as quantum optics, semiclassical quantization of systems 
with spin degrees of freedom, construction of quantum mechanical path 
integrals, the geometric quantization of coadjoint orbits, and so on. One 
specific motivation
of our study lies in the realm of noncommutative spaces. It has been recently 
observed~\cite{AC07} that the fuzzy torus can be regarded as a single mode  
$q$-deformed parafermion. It is important to investigate higher order 
generalizations and supersymmetrization of such fuzzy spaces and their 
relations with $n$-body parastatistics. In this context the coherent states are
expected to provide the star product structure that reflects the 
noncommutativity of such spaces, and allows building of quantum mechanical and 
field theoretical models on such spaces. Coherent state realization~\cite{APS01} 
of such star product, and its generalizations for noncommutative 
superspaces utilizing, for instance,  the oscillator representation of 
$\osp(1|2)$ coherent states have been
achieved~\cite{BKR02}. From these considerations the construction of the 
$\osp(1|2 n)$ coherent states based on $n$-body paraboson representations 
may have utility in analyzing fuzzy superspaces. Another motivation of 
the present work stems from the study of integrable models in quantum mechanics
such as the Calogero model~\cite{C69}. It has been observed~\cite{BEM93},
\cite{M94} that the single mode $(n=1)$ paraboson play a crucial role in 
understanding the Calogero model \cite{C69}. Our construction of the coherent 
states of the multi-mode parabosons, therefore, may facilitate a detailed 
construction of related quantum integrable models.      

The plan of the present work is as follows. In section 2, 
we define the Lie superalgebra $\osp(1|4)$ and give a description of its 
paraboson Fock representations $V(p)$. The $b_1^-$-coherent states are constructed in 
section 3. Since the operators $b_1^{-}$ and $b_2^-$ do not commute
but $b_1^-$ and $(b_2^-)^2$ do, in section 4 we are dealing with the $b_1^-$ and $(b_2^-)^2$-coherent states. 
Section 5 is devoted to the $b_2^-$-matrix elements. The constructed coherent states allow the resolution
of unity which is considered in section 6. We end the paper with some final remarks.

\setcounter{equation}{0}
\section{The Lie superalgebra $\osp(1|4)$ and its paraboson representations} \label{sec:osp}%

In  matrix form the orthosymplectic 
Lie superalgebra $\osp(1|4)$~\cite{Kac} can be defined as a subset of all five by five matrices 
\begin{equation}
\left(\begin{array}{ccccc} 
0&a&b&c&d\\
c&e&f&x&y\\
d&g&h&y&z\\
-a&u&v&-e&-g\\
-b&v&w&-f&-h
\end{array}\right),
\label{osp14}
\end{equation}
where the nonzero entries are arbitrary complex numbers. The even subalgebra
$\osp(1|4)_{{\bar{0}}}$ consists of the matrices for which $a=b=c=d=0$ and the odd
subspace $\osp(1|4)_{{\bar{1}}}$ of $\osp(1|4)$ corresponds to the case in~(\ref{osp14})
when $e=f=g=h=x=y=z=u=v=w=0$.
Let the row and column indices run from $0$ to $4$ 
and  denote by $e_{ij}$ the matrix with zeros everywhere except
a $1$ on position $(i,j)$. Then the Cartan subalgebra $\h$ of $\osp(1|4)$ is 
spanned by the diagonal elements
\begin{equation}
h_1 = e_{11}-e_{33}, \qquad h_2 = e_{22}-e_{44}.
\label{h_j}
\end{equation}
In terms of the dual basis $\delta_1, \; \delta_2$ of $\h^*$, 
the odd root vectors and corresponding roots of $\osp(1|4)$ are given by:
\begin{align}
& e_{0,k}-e_{k+2,0}  \leftrightarrow -\delta_k, \qquad k=1,2, \nn\\
& e_{0,k+2}+e_{k,0}  \leftrightarrow \delta_k, \qquad k=1,2. \nn
\end{align}
The even root vectors  and roots are
\begin{align}
& e_{j,k}-e_{k+2,j+2}  \leftrightarrow \delta_j -\delta_k,  
\qquad j\neq k=1,2,\nn\\
& e_{j,k+2}+e_{k,j+2}  \leftrightarrow  \delta_j +\delta_k, \qquad j\leq k=1,2,\nn\\
& e_{j+2,k}+e_{k+2,j}  \leftrightarrow -\delta_j -\delta_k, \qquad j\leq k=1,2. \nn
\end{align}
We introduce the following multiples of the odd root vectors 
\begin{equation}
b_{k}^+= \sqrt{2}(e_{0, k+2}+e_{k,0}), \qquad
b_{k}^-= \sqrt{2}(e_{0, k}-e_{k+2,0}) \qquad k=1,2.
\label{b-as-e}
\end{equation}
Since all even root vectors can be obtained by anticommutators $\{b_j^{\xi}, b_{ k}^{\eta}\}$,
the following holds~\cite{Ganchev}
\begin{theo}[Ganchev and Palev]
As a Lie superalgebra defined by generators and relations, 
$\osp(1|4)$ is generated by $4$ odd elements $b_k^\pm$ subject to the following  relations:
\begin{equation}
 [\{ b_{ j}^{\xi}, b_{ k}^{\eta}\} , b_{l}^{\epsilon}]= (\epsilon -\xi) \delta_{jl} b_{k}^{\eta} 
 +  (\epsilon -\eta) \delta_{kl}b_{j}^{\xi}.
\label{pboson}
\end{equation}
\end{theo}
The operators $b_j^+$ are the positive odd root vectors, and the
$b_j^-$ are the negative odd root vectors.
Relations~(\ref{pboson}) are the defining triple relations of the paraboson operators.

The paraboson Fock space $V(p)$ is characterized by $(j,k=1,2)$
\begin{equation}
 (b_j^\pm)^\dagger = b_j^\mp, \qquad b_j^- |0\rangle = 0, \qquad
 \{b_j^-,b_k^+\} |0\rangle = p\,\delta_{jk}\, |0\rangle.
 \label{pFock}
\end{equation}
Furthermore, it is easy to verify that
\begin{equation}
\{b_j^-,b_j^+\}=2 h_j \qquad(j=1,2).
\label{bbh}
\end{equation}
Hence we have the following:
\begin{coro}
The paraboson Fock space $V(p)$ is the unitary irreducible representation (unirrep) of
$\osp(1|4)$ with lowest weight $(\frac{p}{2}, \frac{p}{2})$.
\end{coro} 
In~\cite{paraboson}, an explicit basis and the matrix elements of these representations were
constructed. We summarize the results:
\begin{theo}[\cite{paraboson}]
The $\osp(1|4)$ representation $V(p)$ with lowest weight $(\frac{p}{2},\frac{p}{2})$
is a unirrep if and only if $p\geq 1$.
For $p>1$,  the representation space is spanned by the following orthonormal basis (Gelfand-Zetlin basis (GZ)):
\begin{equation}
|m) =  
\left| \begin{array}{l} m_{12},m_{22} \\ m_{11} \end{array} \right),
\label{m2}
\end{equation}
where $(m_{12},m_{22})$ is a partition $\lambda$ of length at most~2, i.e.
$m_{12}$ and $m_{22}$ are integers with 
\begin{equation}
m_{12} \geq m_{22} \geq 0, \quad
and \quad m_{12}\geq m_{11}\geq m_{22}.
\label{betweenness}
\end{equation}
For $p=1$, the basis consists of all vectors~(\ref{m2})-(\ref{betweenness})
with $m_{22}=0$.
The explicit action of the $\osp(1|4)$ generators in $V(p)$ is given by:
\begin{align}
b_1^+ | m )  =& \sqrt{m_{11}-m_{22}+1}\; f_1(m_{12},m_{22}) 
\left| \begin{array}{l} m_{12}+1, m_{22} \\ m_{11}+1 \end{array} \right) \nn\\
 & -\sqrt{m_{12}-m_{11}}\; f_2(m_{12},m_{22}) 
\left| \begin{array}{l} m_{12},m_{22}+1 \\ m_{11}+1 \end{array} \right),\label{sol1}\\
b_2^+ | m )  =& \sqrt{m_{12}-m_{11}+1}\; f_1(m_{12},m_{22}) 
\left| \begin{array}{l} m_{12}+1, m_{22} \\ m_{11} \end{array} \right) \nn\\
 & +\sqrt{m_{11}-m_{22}}\; f_2(m_{12},m_{22}) 
\left| \begin{array}{l} m_{12},m_{22}+1 \\ m_{11} \end{array} \right),\label{sol2}\\
b_1^- | m )  =& \sqrt{m_{11}-m_{22}}\; f_1(m_{12}-1,m_{22}) 
\left| \begin{array}{l} m_{12}-1, m_{22} \\ m_{11}-1 \end{array} \right) \nn\\
 & -\sqrt{m_{12}-m_{11}+1}\; f_2(m_{12},m_{22}-1) 
\left| \begin{array}{l} m_{12},m_{22}-1 \\ m_{11}-1 \end{array} \right),\label{sol3}\\
b_2^- | m )  =& \sqrt{m_{12}-m_{11}}\; f_1(m_{12}-1,m_{22}) 
\left| \begin{array}{l} m_{12}-1, m_{22} \\ m_{11} \end{array} \right) \nn\\
 & +\sqrt{m_{11}-m_{22}+1}\; f_2(m_{12},m_{22}-1) 
\left| \begin{array}{l} m_{12},m_{22}-1 \\ m_{11} \end{array} \right), \label{sol4}\\
h_1 | m )  =& (\frac{p}{2}+m_{11}) | m ), \quad h_2 | m )  = (\frac{p}{2}+m_{12}+m_{22}-m_{11}) | m )
 \label{hi}
\end{align}
where
\begin{align}
f_1(m_{12},m_{22}) &= (-1)^{m_{22}} \frac{(m_{12}+2+{\cal E}_{m_{12}}(p-2))^{1/2}}
{(m_{12}-m_{22}+1+{\cal O}_{m_{12}-m_{22}})^{1/2}}, \\
f_2(m_{12},m_{22}) &= \frac{(m_{22}+1+{\cal E}_{m_{22}}(p-2))^{1/2}}
{(m_{12}-m_{22}+1-{\cal O}_{m_{12}-m_{22}})^{1/2}}
\end{align}
and the {\em even} and {\em odd functions} ${\cal E}_j$ and ${\cal O}_j$ are defined by
\begin{align}
& {\cal E}_{j}=1 \hbox{ if } j \hbox{ is even and 0 otherwise},\nn\\
& {\cal O}_{j}=1 \hbox{ if } j \hbox{ is odd and 0 otherwise}. \label{EO}
\end{align}
\end{theo}
In the following sections, we shall assume that we are dealing with the generic case $p>1$.

\setcounter{equation}{0}
\section{$b_1^-$-coherent states} \label{sec:b1m}%

In this section we will construct coherent states of the operator $b_1^-$ as eigenstates  
in $V(p)$
\begin{equation}
 b_1^-\psi =\alpha \psi,
 \label{cohe}
\end{equation}
where $\alpha$ is a complex eigenvalue. Let $|\zeta \rangle \in V(p)$ be a weight vector annihilated 
by $b_1^-$, i.e.
\begin{equation}
 h_1|\zeta \rangle  =\zeta_1 |\zeta \rangle , \;\; h_2|\zeta \rangle  =\zeta_2 |\zeta \rangle, \;\; 
 b_1^-|\zeta \rangle =0.
 \label{zero}
\end{equation}

\begin{lemm}
Let $|\zeta \rangle \in V(p)$ be a weight vector annihilated 
by $b_1^-$ and let $T_1=b_1^-b_1^+ \in U(\osp(1|4))$.
Then:
\begin{align}
\bullet\quad &
b_1^-(b_1^+)^n|\zeta\rangle=(n+{\cal O}_n(2\zeta_1-1))(b_1^+)^{n-1}|\zeta \rangle
 \label{b1m}\\
\bullet\quad & T_1(b_1^+)^n|\zeta\rangle=(n+1+{\cal E}_n(2\zeta_1-1))(b_1^+)^{n}|\zeta \rangle  \label{T1}
\end{align}
For vectors $v$ in $V(p)$ which are $T_1$-eigenvectors with non-zero eigenvalue, i.e. $T_1v=\lambda v$,
we define $T_1^{-1}v=\lambda^{-1}v$. Then:
\begin{align}
\bullet\quad & T_1^{-1}(b_1^+)^n|\zeta\rangle=(n+1+{\cal E}_n(2\zeta_1-1))^{-1}(b_1^+)^{n}|\zeta \rangle \label{T1m1}\\
\bullet\quad &
(b_1^+T_1^{-1})^n|\zeta\rangle=\prod_{k=1}^n(k+{\cal O}_k(2\zeta_1-1))^{-1}(b_1^+)^{n}|\zeta \rangle
 \label{b1T1}
 \end{align}
\end{lemm} 
\noindent {\bf Proof.} Equation~(\ref{b1m}) holds for $n=1$: 
\[ b_1^-b_1^+|\zeta\rangle=\{ b_1^-,b_1^+\}
|\zeta \rangle= 2h_1|\zeta \rangle =2\zeta_1|\zeta\rangle
\]
 and for $n=2$:
\[ b_1^-(b_1^+)^2|\zeta\rangle
=[b_1^-,(b_1^+)^2] |\zeta \rangle =2b_1^+|\zeta\rangle.
\] 
In the last expression we used  the triple relation 
$[b_1^-,(b_1^+)^2]=2b_1^+$ (see~(\ref{pboson})). Now the result follows using induction on $n$: 
\begin{align}
b_1^-(b_1^+)^n|\zeta\rangle
&=
[ b_1^-, (b_1^+)^2(b_1^+)^{n-2}]|\zeta\rangle \nn\\
& =[ b_1^-, (b_1^+)^2](b_1^+)^{n-2}|\zeta\rangle+(b_1^+)^2[b_1^-, (b_1^+)^{n-2}]|\zeta\rangle
\nn \\
&
=(2(b_1^+)^{n-1}+(b_1^+)^2(n-2+{\cal O}_n(2\zeta_1-1))(b_1^+)^{n-3})|\zeta \rangle
 \nn \\
& =(n+{\cal O}_n(2\zeta_1-1))(b_1^+)^{n-1})|\zeta \rangle.
\nn 
\end{align}
Formula~(\ref{T1}) follows directly from~(\ref{b1m}). Because of the diagonal action of $T_1$
on weight vectors $(b_1^+)^n|\zeta \rangle$ of $V(p)$ and the fact that $n+1+{\cal O}_n(2\zeta_1-1)>0$ one 
concludes that~(\ref{T1m1}) holds. 
Note that $T_1^{-1}$ is not an element of the enveloping algebra; nevertheless its action
on such vectors of $V(p)$ is well defined.
The proof of~(\ref{b1T1}) uses~(\ref{T1m1}) and again induction.

\hskip 16 cm $\Box$

The last result allows us to define a ``vertex operator'' $\chi(\alpha)$:
\begin{equation}
\chi(\alpha)=\sum_{n=0}^\infty \alpha^n(b_1^+T_1^{-1})^n=\frac{1}{1-\alpha b_1^+T_1^{-1}}
\label{vertex}
\end{equation} 
on     vectors $ |\zeta\rangle$ of the form~(\ref{zero}). Then we have:  
\begin{lemm} 
Let $ |\zeta\rangle\in V(p)$ be a weight vector annihilated by $b_1^-$ that is
normalized (i.e. $\langle \zeta | \zeta \rangle=1$), and $\chi(\alpha)$ a vertex
operator of the form~(\ref{vertex}). Then:
\begin{itemize}
\item
$ \chi(\alpha)|\zeta\rangle\in V(p)$
\item
The norm of $\chi(\alpha)|\zeta\rangle$  is  given  by
\begin{equation}
 \langle \;\chi(\alpha)|\zeta\rangle\; | \;\chi(\alpha)|\zeta\rangle \;\rangle = 
{}_0F_1\left({- \atop \zeta_1};\Big(\frac{\bar\alpha \alpha }{2}\Big)^2\right) +
\frac{\bar\alpha \alpha}{2\zeta_1}{\;}_0F_1\left({- \atop \zeta_1+1};\Big(\frac{\bar\alpha \alpha }{2}\Big)^2\right),\label{norm} 
\end{equation}
where ${}_0F_1\left({- \atop a};x\right)$  is  the  classical  hypergeometric series 
\begin{equation}
{}_0F_1\left({- \atop a};x\right)=\sum_{k=0}^\infty \frac{x^k}{(a)_kk!},
\quad
(a)_k=a(a+1)\cdots (a+k-1)
\label{hypergeometric}
\end{equation}
\item $ \chi(\alpha)|\zeta\rangle$ is  an  eigenvector of $ b_1^-$ with 
eigenvalue  $\alpha$:  
\begin{equation}
b_1^-\chi(\alpha)|\zeta\rangle= \alpha \chi(\alpha)|\zeta\rangle.
\label{CSs}
\end{equation}
\end{itemize}
\end{lemm} 
\noindent {\bf Proof.} The first assertion follows from~(\ref{norm}), since it is sufficient to show that the norm 
of the vector is finite. Since vectors of different weights are orthogonal, one has 
\begin{align}
&
\langle \;\chi(\alpha)|\zeta\rangle\; | \; \chi(\alpha)|\zeta\rangle \;\rangle = \sum_{n=0}^\infty \bar{\alpha}^n
\alpha^n \langle \; (b_1^+T_1^{-1})^n |\zeta\rangle\; | \; (b_1^+ T_1^{-1})^n|\zeta\rangle \; \rangle.
\nn 
\end{align}
Consider 
\begin{align}
& \langle \; (b_1^+T_1^{-1})^{n+1} |\zeta\rangle \; | \;
(b_1^+ T_1^{-1})^{n+1}|\zeta\rangle \; \rangle
\nn \\
& =\langle \; b_1^+T_1^{-1}(b_1^+T_1^{-1})^{n} |\zeta\rangle \; | \; 
b_1^+T_1^{-1} (b_1^+ T_1^{-1})^{n}|\zeta\rangle\; \rangle
\nn \\
& =\langle \; T_1^{-1}(b_1^+T_1^{-1})^{n} |\zeta\rangle \; | \;
b_1^-b_1^+T_1^{-1} (b_1^+ T_1^{-1})^{n}|\zeta\rangle \; \rangle\nn \\
& =\langle \; T_1^{-1}(b_1^+T_1^{-1})^{n} |\zeta\rangle \; | \;
(b_1^+ T_1^{-1})^{n}|\zeta\rangle \; \rangle\nn \\ 
& = (n+1+{\cal{E}}_n(2\zeta_1-1))^{-1}
\langle \; (b_1^+T_1^{-1})^{n} |\zeta\rangle \; | \;
(b_1^+ T_1^{-1})^{n}|\zeta\rangle \; \rangle
.
\nn 
\end{align}
In the last expression we used~(\ref{b1m}) and 
(\ref{T1m1}). Now by induction it follows that 
\begin{equation}
 \langle \; (b_1^+T_1^{-1})^{n} |\zeta\rangle \; | \;
(b_1^+ T_1^{-1})^{n}|\zeta\rangle \; \rangle
=\prod_{k=1}^n(k+{\cal{O}}_k(2\zeta_1-1))^{-1}.
\end{equation}
Therefore 
\begin{align} 
&
\langle \;\chi(\alpha)|\zeta\rangle \; | \; \chi(\alpha)|\zeta\rangle \;\rangle = \sum_{n=0}^\infty \bar{\alpha}^n
\alpha^n \prod_{k=1}^n(k+{\cal{O}}_k(2\zeta_1-1))^{-1}
 \nn\\
& =1+\frac{(\frac{\bar\alpha \alpha}{2})^2}{(\zeta_1)1!}+\frac{(\frac{\bar\alpha \alpha}{2})^4}{(\zeta_1)(\zeta_1+1)2!}
+\cdots
\nn \\
&
+\frac{\bar\alpha \alpha}{2\zeta_1}\Big( 1+\frac{(\frac{\bar\alpha \alpha}{2})^2}{(\zeta_1+1)1!}+\frac{(\frac{\bar\alpha \alpha}{2})^4}{(\zeta_1+1)(\zeta_1+2)2!}
+\cdots\Big)
 \nn \\
& ={}_0F_1\left({- \atop \zeta_1};\Big(\frac{\bar\alpha \alpha }{2}\Big)^2\right) +
\frac{\bar\alpha \alpha}{2\zeta_1}{\;}_0F_1\left({- \atop \zeta_1+1};\Big(\frac{\bar\alpha \alpha }{2}\Big)^2\right).
\nn 
\end{align}
 Since the classical hypergeometric series~(\ref{hypergeometric}) is convergent for 
any $x$ one concludes $\chi(\alpha)|\zeta\rangle\in V(p)$.

The last part follows from the following computation:
\begin{align} 
&
b_1^-\chi(\alpha)|\zeta\rangle  
 \nn\\
& =b_1^-\big( 1+\alpha b_1^+T_1^{-1}+\alpha^2 (b_1^+T_1^{-1})(b_1^+T_1^{-1})+
\cdots\big) |\zeta\rangle
\nn \\
&
=\big(b_1^-+\alpha T_1T_1^{-1}+\alpha^2 (T_1T_1^{-1})(b_1^+T_1^{-1})+
\cdots\big) |\zeta\rangle
 \nn \\
& =b_1^-|\zeta \rangle+\alpha \big( 1+ \alpha( b_1^+T_1^{-1})+\alpha^2 (b_1^+T_1^{-1})^2
+\cdots \big) |\zeta \rangle=\alpha \chi(\alpha)|\zeta\rangle.
\nn 
\end{align}

\hskip 16 cm $\Box$

The above considerations show that in order to construct $b_1^-$-coherent states we must find a complete basis 
of the subspace of weight vectors of $V(p)$, annihilated by $b_1^-$. The weight of the vector 
$|m)$ is given by
$
(\frac{p}{2},\frac{p}{2}) + (m_{11},m_{12}+m_{22}-m_{11})
$ (see~(\ref{hi})). Now
if we consider the weights, one could construct vectors
\begin{equation}
|\zeta_{jk}\rangle =\sum_{i=0}^jc_i(j,k)
\left|
\begin{array}{l}
 k+i,j-i  \\  j  \end{array}\right), \; k=0,1,\cdots,\;\;j=0,1,\cdots,k,
\label{zeromodes}
\end{equation}
of weight $(\frac{p}{2},\frac{p}{2}) + (j,k)$ with $b_1^-|\zeta_{jk}\rangle=0$
and $\langle \zeta_{jk}|\zeta_{jk} \rangle =1$ (in other words $\sum_{i=0}^jc_i(j,k)^2=1$).
This construction is given by:
\begin{prop}
An orthonormal basis of the subspace of weight vectors of $V(p)$,
annihilated by $b_1^-$ is given by~(\ref{zeromodes}),
where
\begin{align}
&  c_i(j,k)=\sqrt{\binom{k-j+i}{i}}
\prod_{r=0}^{k-j}\sqrt{\frac{r+1+{\cal O}_r(p-2+2j)}{k+1-r+{\cal O}_{k-r}(p-2)}} \nn \\
& \times \prod_{s=1}^i(-1)^{j-s}\sqrt{\frac{(j+1-s+{\cal E}_{j-s}(p-2))(k-j+2s+{\cal O}_{k+j-1})}
{(k+1+s+{\cal E}_{k+s-1}(p-2))(k-j+2s-{\cal O}_{k+j-1})}}.
\label{coef}
\end{align}
\end{prop}
\noindent {\bf Proof.}
The action of  $b_1^-$ on the GZ basis vectors gives 
\begin{align} 
&
b_1^-|\zeta_{jk}\rangle =
\sum_{i=0}^{j-1}\Big( c_{i+1}(j,k)\sqrt{i+1}f_1(k+i,j-i-1)-
c_i(j,k)\sqrt{k+i-j+1}f_2(k+i,j-i-1)\Big) \nn \\
& \hskip 10 cm \times \left| \begin{array}{l} k+i,j-i-1 \\ j-1 \end{array} \right).
 \nn
\end{align}
Therefore
\begin{align} 
&
c_{i+1}(j,k)=(-1)^{j-i-1}\sqrt{\frac{(k+i-j+1)(j-i+{\cal{E}}_{j-i-1}(p-2))(k-j+2i+2+{\cal{O}}_{k+j-1})}
{(i+1)(k+i+2+{\cal{E}}_{k+i}(p-2))(k-j+2i+2-{\cal{O}}_{k+j-1})}} c_i(j,k). \nn 
\end{align}
Clearly, the coefficients  $c_i(j,k)$ (see~(\ref{coef})) satisfy the last equation.
The condition $\sum_{i=0}^jc_i(j,k)^2=1$ is equivalent to  the following identity:
\begin{align} 
&
\sum_{i=0}^j{k-j+i\choose i}
\prod_{r=1}^i \frac{(j+1-r+{\cal{E}}_{j-r}(p-2))(k-j+2r+{\cal{O}}_{k+j-1})}
{(k+1+r+{\cal{E}}_{k+r-1}(p-2))(k-j+2r-{\cal{O}}_{k+j-1})}\nn \\
& \hskip 6.5cm =\prod_{r=0}^{k-j}  \frac{(k+1-r+{\cal{O}}_{k-r}(p-2))}
{(r+1+{\cal{O}}_{r}(p-2+2j))} .    
 \label{id}
\end{align}
We only sketch the proof of this identity. One has to consider four cases with 
$j$ and $k$ even or odd. In each of these cases the proof follows the following steps:

1) Consider the sum over $i$ even and over $i$ odd separately.

2) Rewrite the sums in hypergeometric form and use a summation theorem.

3) Combine and see that this is the right-hand side of~(\ref{id}).

To show that vectors of the form~(\ref{zeromodes}) form a basis of the space annihilated by 
$b_1^-$, one uses a weight argument and the explicit action of $b_1^-$, given by~(\ref{sol3}).
Note that in $V(p)$, the multiplicity of the weight $(\frac{p}{2}, \frac{p}{2})+(j,k)$ 
is given by $\min (j+1,k+1)$. For $k=0$, it follows from~(\ref{sol3}) that there is only one vector annihilated 
by $b_1^-$. For $k=1$, (\ref{sol3}) and the above multiplicity allow the construction of only two vectors
annihilated by $b_1^-$. More generally, the multiplicity argument and (\ref{sol3}) yield at
most $k+1$ vectors annihilated by $b_1^-$ for a given $k$-value. Since all vectors~(\ref{zeromodes}) 
are linearly independent, the statement follows.

\hskip 16 cm $\Box$

Combination of Proposition 6 and the previous lemma now yields the 
following result:
\begin{prop}
A complete (actually, an overcomplete) set of $b_1^-$-coherent states
$b_1^-\tilde\psi_{jk}(\alpha) = \alpha \tilde\psi_{jk}(\alpha)$
is defined by
\begin{equation}
\tilde\psi_{jk}(\alpha)=\chi(\alpha)|\zeta_{jk}\rangle , \quad k=0,1,\cdots; \;j=0,1,\cdots,k,
\label{coher}
\end{equation}
where $\chi(\alpha)$ and $|\zeta_{jk}\rangle $ are given by~(\ref{vertex}) and~(\ref{zeromodes})-(\ref{coef}) resp. and
\begin{equation}
\langle \;\tilde\psi_{jk}(\alpha)\; | \; \tilde\psi_{jk}(\alpha)\rangle = 
{}_0F_1\left({- \atop \frac{p}{2}+j};\Big(\frac{\bar\alpha \alpha }{2}\Big)^2\right) +
\frac{\bar\alpha \alpha}{p+2j}{\;}_0F_1\left({- \atop \frac{p}{2}+j+1};\Big(\frac{\bar\alpha \alpha }{2}\Big)^2\right)
\label{normcoher}
\end{equation}
\end{prop}

\noindent {\bf Proof.} The only part left to be proved is~(\ref{normcoher}). It follows directly from the fact that 
$
T_1|\zeta_{jk}\rangle=2h_1|\zeta_{jk}\rangle =(p+2j)|\zeta_{jk}\rangle  
$
and~(\ref{norm}).

\hskip 16 cm $\Box$

In a later section, it will be convenient to have this norm~\eqref{normcoher} expressed in
a different way, using the modified Bessel function
\begin{equation}
I_{\nu}(x) = \sum_{n = 0}^{\infty}\frac{(x/2)^{\nu + 2 n}}{n!\,
\Gamma(\nu + n + 1)}.
\label{Bessel}
\end{equation}
Comparing with~\eqref{hypergeometric} implies that~\eqref{normcoher} can be rewritten as
\begin{equation}
{\cal N}_{p,j,\alpha} = \langle \;\tilde\psi_{jk}(\alpha)\; | \; \tilde\psi_{jk}(\alpha)\rangle = 
\left( \frac{\bar\alpha \alpha}{2} \right)^{1-j-p/2} \Gamma(\frac{p}{2}+j)
\left( I_{j-1+p/2}(\bar\alpha \alpha) + I_{j+p/2} (\bar\alpha \alpha) \right).
\label{normcoher2}
\end{equation}

\setcounter{equation}{0}
\section{$b_1^-$- and $(b_2^-)^2$-coherent states} \label{sec:b1mb2m2}%

Using the defining triple paraboson relations~(\ref{pboson}) it is straightforward 
to see that the operators $(b_2^\pm)^2$ commute with 
$b_1^-$ and $b_1^+$.  Hence, the action of $(b_2^\pm)^2$ also commutes with  $T_1$ and $T_1^{-1}$. 
Therefore one concludes 
that $(b_2^\pm)^2$ commutes with $\chi(\alpha)$:
\begin{equation}
(b_2^\pm)^2\tilde\psi_{jk}(\alpha)=\chi(\alpha)(b_2^\pm)^2|\zeta_{jk}\rangle .
\end{equation}
First note that $(b_2^-)^2|\zeta_{jk}\rangle$ is a vector of weight $(\frac{p}{2}, \frac{p}{2})+(j,k-2)$ and
second $b_1^-(b_2^-)^2|\zeta_{jk}\rangle=(b_2^-)^2 b_1^-|\zeta_{jk}\rangle=0$. Since there is only one vector 
of weight  $(\frac{p}{2}+j, \frac{p}{2}+k-2)$ annihilated by $b_1^-$ one concludes that 
$(b_2^-)^2|\zeta_{jk}\rangle=c |\zeta_{j,k-2}\rangle$. We could find the  constant $c$ by computing
$(b_2^-)^2|\zeta_{jk}\rangle $ on one of the GZ basis vectors of $|\zeta_{jk}\rangle$ and compare the result with the same GZ vector in 
$|\zeta_{j,k-2}\rangle$. The result follows
\begin{equation}
(b_2^-)^2|\zeta_{jk}\rangle=\sqrt{(k-1-j+{\cal{E}}_{k-j})(p+k-2+j+{\cal{O}}_{k+j})}|\zeta_{j,k-2}\rangle.
\end{equation}  
In a similar way one obtains 
\begin{equation}
(b_2^+)^2|\zeta_{jk}\rangle=\sqrt{(k+1-j+{\cal{E}}_{k-j})(p+k+j+{\cal{O}}_{k+j})}|\zeta_{j,k+2}\rangle.
\end{equation}  
Therefore
\begin{align} 
&
(b_2^-)^2\tilde\psi_{jk}(\alpha)=\sqrt{(k-1-j+{\cal{E}}_{k-j})(p+k-2+j+{\cal{O}}_{k+j})}\tilde\psi_{j,k-2}(\alpha)
 \label{b2m2psi}\\
& (b_2^+)^2\tilde\psi_{jk}(\alpha)=\sqrt{(k+1-j+{\cal{E}}_{k-j})(p+k+j+{\cal{O}}_{k+j})}\tilde\psi_{j,k+2}(\alpha)
\label{b2+2psi}.
\end{align}
Now it is not difficult to construct  the {\it bicoherent} states which are 
common eigenstates of the mutually commuting $b_1^-$ and $(b_2^-)^2$ operators:
\begin{align} 
&
b_1^-\Psi_{jl}(\alpha,\beta)=\alpha \Psi_{jl}(\alpha,\beta), \quad 
(b_2^-)^2\Psi_{jl}(\alpha,\beta)=\beta \Psi_{jl}(\alpha,\beta),
 \label{b1mb2m2}
\end{align} 
where 
\begin{align}
&\Psi_{jl}(\alpha,\beta) =  \sum_{k=0}^\infty \frac{\beta^{k+\lfloor \frac{l}{2} \rfloor}}{\sqrt{(2k)!!(p+2l)(p+2l+2)\cdots (p+2l+2(k-1))}}\tilde\psi_{j,2k+l}(\alpha),  \label{Psi} \\ 
 &{\hskip 8cm j=0,1,\ldots; \;\;l=j,j+1}.\nn
\end{align}
The index $jl$ in $\Psi_{jl}$ refers to the weight of the lowest weight vector 
in the expansion of 
$\Psi_{jl}$ in the GZ-basis (just as this was the case for $\tilde\psi_{jk}$). We 
will make a brief comment about certain potential utilities of the bicoherent 
states (\ref{Psi}) in the Conclusion.   

\setcounter{equation}{0}
\section{$b_2^-$-matrix elements} \label{sec:b2m}%

In the previous section,~(\ref{b2m2psi})  yields the matrix elements of $(b_2^-)^2$ for 
the set of $b_1^-$-coherent states $\tilde\psi_{jk}(\alpha)$. The matrix elements of $b_2^-$
for this set can also be computed.
Let us consider the operator $\chi(\alpha)$ acting on a weight vector $|\zeta\rangle$ annihilated by $b_1^-$, 
and apply formula~(\ref{b1T1}). Then one could formally write 
\begin{align} 
\chi(\alpha)|\zeta\rangle &=\sum_{n=0}^\infty \alpha^n(b_1^+T_1^{-1})^n|\zeta\rangle \nn\\
& =\sum_{n=0}^\infty \alpha^{2n}(b_1^+T_1^{-1})^{2n}|\zeta\rangle +
\sum_{n=0}^\infty \alpha^{2n+1}(b_1^+T_1^{-1})^{2n+1}|\zeta\rangle\nn\\
& =\sum_{n=0}^\infty \frac{1}{ n!(\zeta_1)_n}\big(\frac{\alpha b_1^+}{2}\big)^{2n}|\zeta\rangle +
\frac{\alpha b_1^+}{2\zeta_1}\sum_{n=0}^\infty \frac{1}{n!(\zeta_1+1)_n}\big(\frac{\alpha b_1^+}{2}\big)^{2n}|\zeta\rangle\nn \\
&
={}_0F_1\left({- \atop \zeta_1};\Big(\frac{\alpha b_1^+ }{2}\Big)^2\right)|\zeta\rangle +
\frac{\alpha b_1^+}{2\zeta_1}{\;}_0F_1\left({- \atop \zeta_1+1};\Big(\frac{\alpha b_1^+ }{2}\Big)^2\right)|\zeta\rangle.
\label{vertex1}
\end{align} 
Now we compute $b_2^-$-matrix elements for the coherent states. Note that, by a weight argument, 
$\langle \tilde\psi_{j', k'}(\alpha ')|b_2^-|\tilde\psi_{jk}(\alpha)\rangle$ can be nonzero only if $k'=k-1$.
First, use~(\ref{vertex1}) and
the fact that $b_2^-$ commutes with $(b_1^+)^2$:
\begin{align} 
& \langle \tilde\psi_{j', k-1}(\alpha ')|b_2^-|\tilde\psi_{jk}(\alpha)\rangle\nn\\
& =\langle \tilde\psi_{j', k-1}(\alpha ')|b_2^-
\left( {}_0F_1\left({- \atop \frac{p}{2}+j};\Big(\frac{\alpha b_1^+ }{2}\Big)^2\right) +
\frac{\alpha b_1^+}{p+2j}{\;}_0F_1\left({- \atop \frac{p}{2}+j+1};\Big(\frac{\alpha b_1^+ }{2}\Big)^2\right)  \right)|\zeta_{jk}\rangle
\nn\\
&=\langle \tilde\psi_{j', k-1}(\alpha ')|
 {\;}_0F_1\left({- \atop \frac{p}{2}+j};\Big(\frac{\alpha b_1^+ }{2}\Big)^2\right) b_2^- |\zeta_{jk}\rangle+
\langle \psi_{j', k-1}(\alpha ')|
{\;}_0F_1\left({- \atop \frac{p}{2}+j+1};\Big(\frac{\alpha b_1^+ }{2}\Big)^2\right)  
\frac{\alpha b_2^-b_1^+}{p+2j}
|\zeta_{jk}\rangle.
\nn
\end{align}
Now, use the action of $b_1^+$ to the left and the action $b_1^-\tilde\psi_{j,k-1}(\alpha ')=\alpha ' \tilde\psi_{j,k-1}(\alpha ')$.
This yields:
\begin{align}
& \langle \tilde\psi_{j', k-1}(\alpha ')|b_2^-|\tilde\psi_{jk}(\alpha)\rangle \nn\\
& = {}_0F_1\left({- \atop \frac{p}{2}+j};\Big(\frac{\alpha \bar{\alpha}' }{2}\Big)^2\right)
\langle \tilde\psi_{j', k-1}(\alpha ')|b_2^-|\zeta_{jk}\rangle+
{}_0F_1\left({- \atop \frac{p}{2}+j+1};\Big(\frac{\alpha \bar{\alpha}' }{2}\Big)^2\right)
\frac{\alpha}{p+2j}\langle \tilde\psi_{j', k-1}(\alpha ')|b_2^-b_1^+|\zeta_{jk}\rangle.\nn 
\end{align}
So the computation is reduced to computing the above two matrix elements. Using the explicit form of 
$|\zeta_{jk}\rangle $, the action of $b_1^+$ and $b_2^-$ on GZ-basis elements, and the expansion of $\tilde\psi_{j',k-1}(\alpha ')$
in terms of GZ-basis elements one finds:
\begin{align}
\langle \tilde\psi_{j', k-1}(\alpha ')|b_2^-|\zeta_{jk}\rangle 
&= \left\{ \begin{array}{llll} 
{\displaystyle{
\frac{p-2}{p-2+2j}\sqrt{k-j+{\cal{O}}_{k-j}(p-1+2j)}}
 } & \hbox{if} \;\; j'=j \\
{}\\
{ \displaystyle{ 2(-1)^{j-1}{\bar{\alpha}}'\frac{\sqrt{j(p-2+j)(p+k+j-1-{\cal{E}}_{k-j})}}{(p+2j-2)^{3/2}}} 
} & \hbox{if} \;\; j'=j-1 \\
{}\\
{0} & \hbox{otherwise} 
\end{array}\right.\nn \\
\nn\\
\langle \tilde\psi_{j', k-1}(\alpha ')|b_2^-b_1^+|\zeta_{jk}\rangle 
&= \left\{ \begin{array}{llll} 
{\displaystyle{
-\bar{\alpha}'\frac{p-2}{p+2j}\sqrt{k-j+{\cal{O}}_{k-j}(p-1+2j)}}
 } & \qquad\;\; \hbox{if} \;\; j'=j \\
{}\\
{ \displaystyle {2(-1)^{j}\sqrt{\frac{(j+1)(p-1+j)(k-j-{\cal{O}}_{k-j})}{(p+2j)}} }
} & \qquad\;\; \hbox{if} \;\; j'=j+1 \\
{}\\
{0} & \qquad\;\hbox{ otherwise} 
\end{array}\right.
\nn
\end{align}
Hence $\langle \tilde\psi_{j', k-1}(\alpha ')|b_2^-|\tilde\psi_{jk}(\alpha)\rangle$ is $0$ for 
$j'\neq j-1,j,j+1$. In the other cases it is given by:
\begin{align}
& \langle \tilde\psi_{j-1, k-1}(\alpha ')|b_2^-|\tilde\psi_{jk}(\alpha)\rangle \nn\\
& = {}_0F_1\left({- \atop \frac{p}{2}+j};\Big(\frac{\alpha \bar{\alpha}' }{2}\Big)^2\right)
2(-1)^{j-1}{\bar{\alpha}}'\frac{\sqrt{j(p-2+j)(p+k+j-1-{\cal{E}}_{k-j})}}{(p+2j-2)^{3/2}},\nn\\
&\nn\\
& \langle \tilde\psi_{j, k-1}(\alpha ')|b_2^-|\tilde\psi_{jk}(\alpha)\rangle \nn\\
& = {}_0F_1\left({- \atop \frac{p}{2}+j};\Big(\frac{\alpha \bar{\alpha}' }{2}\Big)^2\right)
\frac{p-2}{p-2+2j}\sqrt{k-j+{\cal{O}}_{k-j}(p-1+2j)}\nn\\
&-{}_0F_1\left({- \atop \frac{p}{2}+j+1};\Big(\frac{\alpha \bar{\alpha}' }{2}\Big)^2\right)
\alpha \bar{\alpha}'\frac{p-2}{(p+2j)^2}\sqrt{k-j+{\cal{O}}_{k-j}(p-1+2j)},\nn
& \nn \\& \nn \\
& \langle \tilde\psi_{j+1, k-1}(\alpha ')|b_2^-|\tilde\psi_{jk}(\alpha)\rangle \nn\\
&={}_0F_1\left({- \atop \frac{p}{2}+j+1};\Big(\frac{\alpha \bar{\alpha}' }{2}\Big)^2\right)
2\alpha (-1)^{j}\frac{\sqrt{(j+1)(p-1+j)(k-j-{\cal{O}}_{k-j})}}{(p+2j)^{3/2}}.\nn
\end{align}

\section{Resolution of the identity  via $|\psi_{jk}(\alpha)\rangle$ states}
\label{sec:identity}
\setcounter{equation}{0}

We now discuss the resolution of the identity operator via the $b_{1}^{-}$-coherent 
states $\tilde\psi_{jk}(\alpha)$. To be precise, we restrict ourselves to the states obtained by repeated actions of $b_{1}^{+}$ on the vector $|\zeta_{jk}\rangle$ ($k=0,1,\ldots$; $j=0,1,\ldots,k$) that is annihilated by $b_{1}^{-}$: 
\begin{equation}
|\zeta_{jk};n\rangle = \frac{(b_{1}^{+})^{n}}
{\sqrt{2^{n}\,\big(\frac{n - {\cal O}_{n}}{2}\big)!\,\big(\frac{p}{2} 
+ j\big)_{\frac{n + {\cal O}_{n}}{2}}}}\,|\zeta_{jk}\rangle,\qquad n = 0, 1, 2,
\ldots.
\label{num_st}
\end{equation}
The action of $b_{1}^{\pm}$ operators on these orthonormal states of weights 
$(\frac{p}{2}+j+n, \frac{p}{2}+k)$ are given by
\begin{align}
b_{1}^{+} |\zeta_{jk}; n\rangle &= \sqrt{(p+2j) {\cal E}_{n} + n + {\cal O}_{n}}\,|\zeta_{jk}; n+1 \rangle, \nn\\
b_{1}^{-} |\zeta_{jk}; n+1\rangle &= \sqrt{(p+2j) {\cal E}_{n} + n + {\cal O}_{n}}\,|\zeta_{jk}; n \rangle.\nn
\end{align}
In the following we consider the subspace 
${\cal V}_{\langle j \rangle, k}$ spanned by all vectors $|\zeta_{jk}; n\rangle$ with $n=0,1,2,\ldots$.
For our discussion of the resolution of unity the state 
vectors $|\zeta_{jk}; n\rangle$ and the subspace ${\cal V}_{\langle j \rangle, k}$ play a vital role. 

Let us now consider the {\em normalized} $b_1^-$-coherent states determined by~\eqref{normcoher} and~\eqref{normcoher2}:
\begin{equation}
|\psi_{jk}(\alpha)\rangle = {\cal N}_{p,j,\alpha}^{-1/2} \; \tilde\psi_{jk}(\alpha) =
{\cal N}_{p,j,\alpha}^{-1/2} \; \chi(\alpha) \, |\zeta_{jk}\rangle.
\end{equation}

Employing~\eqref{vertex1} and~\eqref{num_st} the normalized coherent states 
$|\psi_{jk}(\alpha)\rangle$ may be expanded in the discrete basis $|\zeta_{jk}; 
n\rangle$ of the subspace ${\cal V}_{\langle j \rangle, k}$ as 
\begin{align}
&|\psi_{jk}(\alpha)\rangle = {\cal N}_{p,j,\alpha}^{-1/2} \;
\left[{}_0F_{1}\left({- \atop \frac{p}{2}+j};\Big(\frac{\alpha b_{1}^{+}}{2}\Big)^{2}\right) 
+ \frac{\alpha b_{1}^{+}}{p+2j}\; {}_0F_{1}\left({- \atop \frac{p}{2}+j+ 1};\Big(\frac{\alpha b_{1}^{+}}{2}\Big)^{2}\right)\right] \;|\zeta_{jk}\rangle\nn\\
&= \left(\Big(\frac{\bar{\alpha} \alpha}{2}\Big)^{1-\frac{p}{2} -j}\,\Big(I_{\frac{p}{2}+j-1}(\bar{\alpha} \alpha) + I_{\frac{p}{2}+j}(\bar{\alpha} \alpha)\Big)\right)^{- 1/2} \nn\\
&\times 
\sum_{n = 0}^{\infty}\Big(\frac{\alpha^{2 n}}{\sqrt{n!\,2^{2 n}\,\Gamma\big(\frac{p}{2} + j + n\big)}}\,|\zeta_{jk}; 2n\rangle + \frac{\alpha^{2 n + 1}}{\sqrt{n!\,2^{2 n + 1}\,\Gamma\big(\frac{p}{2} + j + n + 1\big)}}\,|\zeta_{jk}; 2n+1\rangle \Big). \label{cost_expn}
\end{align}
Being an overcomplete set the distinct normalized coherent states are not 
orthogonal:
\begin{equation}
\langle\psi_{jk}(\alpha')|\psi_{jk}(\alpha)\rangle =
\frac{\Big(\frac{\bar{\alpha'} \alpha}{2}\Big)^{1-\frac{p}{2} -j}\Big(I_{\frac{p}{2}+j-1}(\bar{\alpha'} \alpha) + I_{\frac{p}{2}+j}(\bar{\alpha'} \alpha)\Big)}{{\Big[\Big(\frac{|\alpha'|^{2}\, |\alpha|^{2}}{4}\Big)^{1-\frac{p}{2} -j}\Big(I_{\frac{p}{2}+j-1}(|\alpha'|^{2}) + I_{\frac{p}{2}+j}(|\alpha'|^{2})\Big)\Big(I_{\frac{p}{2}+j-1}(|\alpha|^{2}) 
+ I_{\frac{p}{2}+j}(|\alpha|^{2})\Big)\Big]^{1/2}}}.
\label{inn_pro}
\end{equation}
The reflection property of the modified Bessel function
$\displaystyle{I_{\nu}(-x) = (-1)^{\nu}\,I_{\nu}(x)}$ requires the 
following inner product to be real:
\begin{equation}
\langle\psi_{jk}(- \alpha)|\psi_{jk}(\alpha)\rangle = \langle\psi_{jk}(\alpha)|\psi_{jk}(- \alpha)\rangle = \frac{I_{\frac{p}{2}+j-1}(|\alpha|^{2}) - I_{\frac{p}{2}+j}(|\alpha|^{2})}{I_{\frac{p}{2}+j-1}(|\alpha|^{2}) + I_{\frac{p}{2}+j}(|\alpha|^{2})}.
\label{real_con}
\end{equation}
We note that the reality of the overlap function~\eqref{real_con} allows us to construct a cat-type two-dimensional subspace with orthogonal bases:
\begin{equation}
|\psi_{jk}(\alpha)\rangle_{\pm} = \frac{|\psi_{jk}(\alpha)\rangle \pm \psi_{jk}(- \alpha)\rangle}{||\psi_{jk}(\alpha)\rangle \pm |\psi_{jk}(- \alpha)\rangle|},
\qquad {}_{+}\langle\psi_{jk}(\alpha)|\psi_{jk}(\alpha)\rangle_{-} = 0,  
\label{alpha_pm}
\end{equation}
where the normalized states $|\psi_{jk}(\alpha)\rangle_{\pm}$ explicitly read 
\begin{align}
& |\psi_{jk}(\alpha)\rangle_{+} = \Big(\Big(\frac{|\alpha|^{2}}{2}\Big)^{1-p/2 -j}\,I_{\frac{p}{2}+j-1}(|\alpha|^{2})\Big)^{- 1/2}\, \sum_{n = 0}^{\infty} \frac{\alpha^{2 n}}{\sqrt{n!\,2^{2 n}\,\Gamma\big(\frac{p}{2} + j + n\big)}}\,
|\zeta_{jk}; 2n\rangle,
\label{alpha_plus} \\
& |\psi_{jk}(\alpha)\rangle_{-} = \Big(\Big(\frac{|\alpha|^{2}}{2}\Big)^{1-\frac{p}{2} -j}\,I_{\frac{p}{2}+j}(|\alpha|^{2})\Big)^{- 1/2}\, \sum_{n = 0}^{\infty} \frac{\alpha^{2 n+ 1}}{\sqrt{n!\,2^{2 n + 1}\,\Gamma\big(\frac{p}{2} + j + n + 1\big)}}\,|\zeta_{jk}; 2n + 1\rangle.
\label{alpha_minus}
\end{align} 

Following ~\cite{KPS01} we now provide a resolution of the identity 
operator on the subspace 
${\cal V}_{\langle j \rangle, k}$ via the coherent states 
$|\psi_{jk}(\alpha)\rangle$. To proceed, using the completeness of the 
discrete orthonormal basis states on the weight space ${\cal V}_{\langle j \rangle, k}$, one can write
\begin{equation}
\sum_{n = 0}^{\infty} |\zeta_{jk}; n\rangle \langle\zeta_{jk}; n| 
= \mathbb {I},
\label{numst_unity}
\end{equation}
where $\mathbb {I}$ stands for the identity operator on ${\cal V}_{\langle j \rangle, k}$.
Using the polar decomposition of the complex plane
\begin{equation}
\alpha = \rho \,\exp (i \theta), \qquad d^{2}\alpha = \frac{\rho\,\,d\rho\,
d\theta}{2 \pi}\nn
\end{equation}
concurrently with our construction (\ref{cost_expn}) of the coherent states  
$|\psi_{jk}(\alpha)\rangle$ we integrate on the angular variable to obtain 
\begin{align}
\int_{0}^{2 \pi}\frac{d\theta}{2 \pi}\,|\psi_{jk}(\alpha)\rangle \langle\psi_{jk}(\alpha)|  
& =\Big(\frac{\rho^{2}}{2}\Big)^{\frac{p}{2} + j - 1}\, \Big(I_{\frac{p}{2}+j-1}(\rho^{2}) + I_{\frac{p}{2}+j}(\rho^{2})\Big)^{- 1}  \nn\\
& \times \sum_{n = 0}^{\infty}\Big(\frac{1}{n!\,\Gamma\big(\frac{p}{2} + j + n\big)}
\,\Big(\frac{\rho^{2}}{2}\Big)^{2n}\,|\zeta_{jk}; 2n\rangle\,\langle\zeta_{jk}; 2n| \nn \\
& + \frac{1}{n!\,\Gamma\big(\frac{p}{2} + j + n + 1\big)}\,\Big(\frac{\rho^{2}}{2}\Big)^{2n+1}\,|\zeta_{jk}; 2n + 1\rangle\,\langle\zeta_{jk}; 2n + 1|\Big).
\label{theta_int}
\end{align}
We observe that in order to construct a resolution of unity on the current
subspace it is necessary to consider the off-diagonal elements 
$|\psi_{jk}(\alpha)\rangle \, \langle\psi_{jk}(- \alpha)|$ of the density matrix. 
Using as yet to be determined functions $F_{I}(\rho),\, F_{II}(\rho)$ over the entire complex $\alpha$
plane, we employ~\eqref{theta_int} to obtain 
\begin{align}
&\int d^{2}\alpha \Big(\frac{\rho^{2}}{2}\Big)^{1 - \frac{p}{2} - j}\, \Big(I_{\frac{p}{2}+j-1}(\rho^{2}) + I_{\frac{p}{2}+j}(\rho^{2})\Big)\, F_{I}(\rho) \times\nn\\
&\qquad \frac{1}{2}\,\Big(|\psi_{jk}(\alpha)\rangle \, \langle\psi_{jk}(\alpha)| + |\psi_{jk}(\alpha)\rangle \, \langle\psi_{jk}(- \alpha)|\Big) =\sum_{n = 0}^{\infty} |\zeta_{jk}; 2n\rangle\,
\langle\zeta_{jk}; 2n|,\label{state_sum_even}\\
&\int d^{2}\alpha \Big(\frac{\rho^{2}}{2}\Big)^{1 - \frac{p}{2} - j}\, \Big(I_{\frac{p}{2}+j-1}(\rho^{2}) + I_{\frac{p}{2}+j}(\rho^{2})\Big)\, F_{II}(\rho) \times\nn\\
&\qquad \frac{1}{2}\,\Big(|\psi_{jk}(\alpha)\rangle \, \langle\psi_{jk}(\alpha)| - |\psi_{jk}(\alpha)\rangle \, \langle\psi_{jk}(- \alpha)|
\Big) = \sum_{n = 0}^{\infty} |\zeta_{jk}; 2n + 1\rangle\,
\langle\zeta_{jk}; 2n + 1|.
\label{state_sum_odd}
\end{align}
The functions  $F_{I}(\rho),\, F_{II}(\rho)$ introduced above are required to 
satisfy the Stieltjes moment relations
\begin{align}
\int_{0}^{\infty} d\rho\,\rho^{4 n + 1}\,F_{I}(\rho) &= 2^{2 n}\,\Gamma(n+1)\,\Gamma\Big(\frac{p}{2} + j + n\Big),\\
\int_{0}^{\infty} d\rho\,\rho^{4 n + 3}\,F_{II}(\rho) &= 2^{2 n + 1}\,\Gamma(n+1)\,\Gamma\Big(\frac{p}{2} + j + n + 1\Big).
\label{F_def}
\end{align}
These functions are now explicitly determined by constructing the inverse 
Mellin transform. Using analytic continuation methods we express them as
\begin{align}
F_{I}(\rho) &= \frac{1}{\pi i}\,\int_{c - i \infty}^{c + i \infty}\,dz \,\Big(\frac{\rho^{2}}{2}\Big)^{- 2 z}\,\Gamma\Big(z + \frac{1}{2}\Big)\,\Gamma\Big(z + \frac{p}{2} + j -\frac{1}{2}\Big), \label{FI}\\
F_{II}(\rho) &= \frac{1}{\pi i}\,\int_{c - i \infty}^{c + i \infty}\,dz \,\Big(\frac{\rho^{2}}{2}\Big)^{- 2 z}\,\Gamma (z)\,\Gamma\Big(z + \frac{p}{2} + 
j\Big),
\label{FII}
\end{align}
where the pole structure of $\Gamma(z)$ on the negative real axis reads
\begin{equation}
\Gamma( - n + \epsilon) = \frac{(- 1)^{n}}{n!}\,\Big(\frac{1}{\epsilon} + \psi(n + 1) + O(\epsilon)\Big), \qquad \psi(z) = (\ln \Gamma(z))'.\label{gam_sing}\\
\end{equation}

The contour integrals listed in~\eqref{FI} and~\eqref{FII} have distinct singularity structures 
for even integral values of $p$, and for the generic case $p > 1$.
We first construct the inverse transforms for the even integral values: 
$p = 2 m$, $m = 1, 2, \ldots$.
Substituting the Laurent expansion~\eqref{gam_sing} in~\eqref{FI} 
it is evident that the integrand in~\eqref{FI} has $m + j - 1$ simple poles at  
$z = - n - 1/2$, $n = 0, 1,\ldots,(m + j - 2)$, and an infinite number of poles of
order $2$ at $z = 1/2 - m - j - n$, $n = 0, 1, 2,\ldots.$ The integral 
vanishes exponentially as $|z| \rightarrow \infty$ on the left half-plane.
Adjoining the contour in~\eqref{FI} with a semicircle $|z| = R$ on the left 
half-plane, and then proceeding to the limiting value of its radius, 
$R \rightarrow \infty$, we get the integral~\eqref{FI} as a sum of the 
contributions arising from the simple and double poles, given, respectively,
as follows: 
\begin{equation}
2\,\sum_{n = 0}^{m + j - 2}\,\frac{(- 1)^{n}}{n!}\,\Big(\frac{\rho^{2}}{2}\Big)^{2 n + 1}\,(m + j - n - 2)!\;,
\label{sing_pol}
\end{equation}
\begin{equation}
2\,(-1)^{m + j - 1}\,\sum_{n = 0}^{\infty}\frac{1}{n!\,(m + j + n - 1)!}\,\Big(\frac{\rho^{2}}{2}\Big)^{p + 2j +2 n - 1}\,\Big(\psi(m + j + n) + \psi(n + 1) -2\,\ln \Big(\frac{\rho^{2}}{2}\Big)\Big).\label{doub_pol}
\end{equation}
Combining the above contributions of the residues we obtain the promised 
explicit expression of the measure: 
\begin{equation}
F_{I}(\rho) = 4 \Big(\frac{\rho^{2}}{2}\Big)^{m + j}\,K_{m +j-1} (\rho^{2}),
\label{FI_K}
\end{equation}
where the modified Bessel function of the second kind $K_{\nu}(z)$ for an 
integral order $\nu$ is given by
\begin{align}
K_{\nu}(z) &= \frac{1}{2}\,\sum_{n = 0}^{\nu - 1} (-1)^{n}\,\frac{(\nu - n - 1)!}{n!}\,
\big(\frac{z}{2}\big)^{2 n - \nu} \nn\\ 
&+ (- 1)^{\nu + 1}\, \sum_{n = 0}^{\infty}\,
\frac{1}{n! (\nu + n)!}\,\Big(\ln \big(\frac{z}{2}\big) - \frac{1}{2} \psi(n + 1) - \frac{1}{2} \psi(\nu + n + 1)\Big).
\label{K_def}
\end{align}
To evaluate the contour integral~\eqref{FII} for the even integral 
values $p = 2\,m$, $m = 1, 2, \ldots$ we notice, as before, that the integrand 
has $m + j$ simple poles at $z = - n$, $n = 0, 1, \ldots, m + j -1$, and an 
infinite number of poles of order $2$ at $z = - m - j - n$, $n = 0, 1, 2, 
\ldots$, respectively. The contribution of the residues at these poles now 
yields the corresponding measure function:
\begin{equation}
F_{II}(\rho) = 4 \Big(\frac{\rho^{2}}{2}\Big)^{m + j}\,K_{m + j} (\rho^{2}).
\label{FII_K}
\end{equation}

We now turn to the case of {\it generic} values of the order parameter $p > 1$.
Contrary to the earlier instance of even integral values of $p$, now the 
contour integral~\eqref{FI} of the inverse Mellin transform has two 
infinite sequences of simple poles at  
$z = - \frac{1}{2} - n$, $n = 0, 1, 2,\ldots$, and $z = \frac{1}{2} - 
\frac{p}{2} - j - n$, $n = 0, 1, 2,\ldots$. The corresponding residues to the 
contour integral~\eqref{FI} read
\begin{align}
& 2\,\sum_{n = 0}^{\infty} \frac{(-1)^{n}}{n!}\,\Big(\frac{\rho^{2}}{2}\Big)^{2 n + 1}\,
 \Gamma\Big(\frac{p}{2} + j - n - 1\Big),\label{gen_pole_1}\\
& 2\,\sum_{n = 0}^{\infty} \frac{(-1)^{n}}{n!}\,\Big(\frac{\rho^{2}}{2}\Big)^{p +2 j + 2 n - 1}\,
 \Gamma\Big(1 - \frac{p}{2} - j - n\Big),
\label{gen_pole_2}
\end{align}
respectively. Employing the reflection property of the gamma functions
\begin{equation}
\Gamma(z)\,\Gamma(1 - z) =\frac{\pi}{\sin \pi z}\nn
\end{equation}
the residues~\eqref{gen_pole_1} and~\eqref{gen_pole_2} may be summed to yield 
the measure  
\begin{align}
F_{I}(\rho) &= \frac{2 \pi}{\sin\, \pi (\frac{p}{2} + j)}\, \Big(\frac{\rho^{2}}{2}\Big)^{\frac{p}{2} + j}\,\Big(I_{\frac{p}{2}+j-1}(\rho^{2}) - I_{1 -\frac{p}{2} - j}(\rho^{2})\Big)\nn\\
&= 4 \Big(\frac{\rho^{2}}{2}\Big)^{\frac{p}{2} + j}
\,K_{\frac{p}{2} +j-1} (\rho^{2}).
\label{FI_gen}
\end{align}
In the second equality in~\eqref{FI_gen} we used the general construction of 
the modified Bessel function of the second kind $K_{\nu}(z)$ of arbitrary 
order $\nu$: 
\begin{equation}
K_{\nu}(z) = \frac{\pi}{2\, \sin (\nu\, \pi)}\,\big(I_{- \nu}(z) 
- I_{\nu}(z)\big).
\label{K_gen}
\end{equation} 
Comparing the measure function~\eqref{FI_gen} with the case~\eqref{FI_K} for 
even integral values of $p$, we observe that the form~\eqref{FI_gen} is valid
for all values of $p > 1$. Proceeding as before we observe that the integrand 
in~\eqref{FII} has two sequences of simple poles at  
$z = - n$, $n = 0, 1, 2,\ldots$, and $z = - \frac{p}{2} - j - n$, 
$n = 0, 1, 2,\ldots$, respectively. Evaluating their contributions we 
produce the measure function $F_{II} (\rho)$:
\begin{align}
F_{II}(\rho) &= \frac{2 \pi}{\sin\, \pi (\frac{p}{2} + j)}\, 
\Big(\frac{\rho^{2}}{2}\Big)^{\frac{p}{2} + j}\,
\Big(I_{- \frac{p}{2} - j}(\rho^{2}) - I_{\frac{p}{2} + j}(\rho^{2})\Big)\nn\\
&= 4 \Big(\frac{\rho^{2}}{2}\Big)^{\frac{p}{2} + j}
\,K_{\frac{p}{2} +j} (\rho^{2}).
\label{FII_gen}
\end{align}
Comparison with~\eqref{FII_K} again reveals that the form~\eqref{FII_gen} of 
the measure is universally true for arbitrary $p > 1$.

The coherent states $|\psi_{jk} (\alpha)\rangle$ now provide a decomposition of the 
identity operator in the subspace ${\cal V}_{\langle j \rangle, k}$ with an
explicitly known weight function. Combining~\eqref{state_sum_even} and~\eqref{FI_gen} 
the projection operator on the even subspace may be realized as 
\begin{align}
&\int d^{2}\alpha \,\rho^{2}\, \Big(I_{\frac{p}{2}+j-1}(\rho^{2}) + I_{\frac{p}{2}+j}(\rho^{2})\Big)\, K_{\frac{p}{2} + j - 1}(\rho^{2}) \Big(|\psi_{jk}(\alpha)\rangle \, \langle\psi_{jk}(\alpha)| + |\psi_{jk}(\alpha)\rangle \, \langle\psi_{jk}(- \alpha)|\Big)\nn\\
&\qquad =\sum_{n = 0}^{\infty} |\zeta_{jk}; 2n\rangle\,\langle\zeta_{jk}; 2n|.
\label{measure_even}
\end{align}
A similar construction of the projection operator on the odd subspace follows
from~\eqref{state_sum_odd} and~\eqref{FII_gen}:
\begin{align}
&\int d^{2}\alpha \,\rho^{2}\, \Big(I_{\frac{p}{2}+j-1}(\rho^{2}) + I_{\frac{p}{2}+j}(\rho^{2})\Big)\, K_{\frac{p}{2} + j} (\rho^{2})\,\Big(|\psi_{jk}(\alpha)\rangle \, \langle\psi_{jk}(\alpha)| - |\psi_{jk}(\alpha)\rangle \, \langle\psi_{jk}(- \alpha)|\Big)\nn\\
&\qquad= \sum_{n = 0}^{\infty} |\zeta_{jk}; 2n + 1\rangle\,\langle\zeta_{jk}; 2n + 1|.
\label{measure_odd}
\end{align}
The decomposition of unity on the subspace ${\cal V}_{\langle j \rangle, k}$ now emerges 
from simultaneous use of~\eqref{measure_even} and~\eqref{measure_odd}:
\begin{align}
&\int d^{2}\alpha \,\rho^{2}\, \Big(I_{\frac{p}{2}+j-1}(\rho^{2}) + I_{\frac{p}{2}+j}(\rho^{2})\Big) \,\Big(\Big(K_{\frac{p}{2} + j - 1} (\rho^{2}) +  K_{\frac{p}{2} + j} (\rho^{2})\Big)\,|\psi_{jk}(\alpha)\rangle \, \langle\psi_{jk}(\alpha)| \nn\\
&\qquad + \Big(K_{\frac{p}{2} + j - 1} (\rho^{2}) -  K_{\frac{p}{2} + j} (\rho^{2})\Big)\,|\psi_{jk}(\alpha)\rangle \, \langle\psi_{jk}(- \alpha)|\Big) = \sum_{n = 0}^{\infty} |\zeta_{jk}; n\rangle \langle\zeta_{jk}; n| = \mathbb{I}.
\label{unit_decom}
\end{align}
As remarked earlier, the decomposition given above includes off-diagonal terms 
of the density operator. The nondiagonal nature of the representation disappears if we express the density operators {\it via} the cat-type $|\psi_{jk}(\alpha)
\rangle_{\pm}$ states introduced in~\eqref{alpha_pm}: 
\begin{align}
&\int\frac{\rho\, d\rho\, d\theta}{\pi}\,\rho^{2}\, \Big(I_{\frac{p}{2}+j-1}(\rho^{2})\, K_{\frac{p}{2} + j - 1} (\rho^{2}) \,|\psi_{jk}(\alpha)\rangle_{+} \,{}_{+}\langle\psi_{jk}(\alpha)| \nn\\
&\qquad\qquad + I_{\frac{p}{2}+j}(\rho^{2})\, K_{\frac{p}{2} + j} (\rho^{2}) \,|\psi_{jk}(\alpha)\rangle_{-} \,{}_{-}\langle\psi_{jk}(\alpha)|\Big) = \mathbb{I}.
\label{dia_decom}
\end{align}
Using the integral representations of the Bessel functions
\begin{align}
I_{\nu}(z) &=\frac{\big(\frac{z}{2}\big)^{\nu}}{\Gamma \big(\nu + 
\frac{1}{2}\big)\; \Gamma\big(\frac{1}{2}\big)}\;\int_{-1}^{1}
(1 - t^{2})^{\nu - \frac{1}{2}}\;\cosh (z t)\, dt, \qquad 
\nu + \frac{1}{2} > 0,\nn\\
K_{\nu}(2 z) &= \int_{0}^{\infty}\;\exp (- z\,\exp t)
\;\exp (- z\,\exp (- t)) \;\cosh (\nu\, t)\;dt 
\label{bessel_int}
\end{align}
we conclude that the weight function of the measure in the decomposition of the 
unity given via the cat-type states in~\eqref{dia_decom} is 
{\it positive definite} for the domain $p > 1$.

\section{Conclusion}
\label{sec:conclusion}

To summerize, we obtained the coherent state representations of the 
$\osp(1|4)$ algebra by constructing the eigenstates of the paraboson 
operator $b_{1}^{-}$. In the subspace ${\cal V}_{\langle j \rangle, k}$
the coherent state vectors $|\psi_{j k}(\alpha)\rangle$ provide a decomposition of unity 
with an explicitly known weight function. When expressed  via the 
cat-type states $|\psi_{j k}(\alpha)\rangle_{\pm}$ this measure assumes a 
positive definite form for the range of the order parameter $p > 1$. In 
addition we have produced  the bicoherent states $\Psi_{jl}(\alpha, \beta)$    
which are eigenstates of the mutually commuting operators 
$b_{1}^{-}$ and $(b_{2}^{-})^{2}$. These states live on the subspace
$\displaystyle{\oplus_{m}\,{\cal V}_{\langle j \rangle, k + 2 m}}$, and their 
completeness on this subspace may be investigated by using the inverse 
Mellin transform method followed here. We hope to discuss this result elsewhere.

We conclude the  paper with certain pointers towards further developments 
along the present lines. It is known that $q$-deformed parafermions play a 
crucial role in understanding the noncommutative space of the fuzzy torus. 
Similarly a $q$-deformed analog of the $n$-mode paraboson algebra may be the
underlying feature of a class of fuzzy superspaces. An extension of the 
coherent states presented here is likely to provide a star product structure for 
such noncommutative superspaces. Lastly, a coordinate representation of the 
trilinear commutation relation of the $n$-mode parabosons is likely to be of
significance. In view of the close affinity of the Calogero model with the 
single mode paraboson, such coordinate representations are likely to enhance 
our understanding of the correspondingly related $n$-body quantum integrable 
Hamiltonian. The bicoherent states and the matrix elements of the 
$b_{2}^{-}$ operator constructed in sections~\ref{sec:b1mb2m2} and~\ref{sec:b2m}, 
respectively, should help us in the description of these 
Hamiltonians.

\section*{Acknowledgments}
N.I.\ Stoilova would like to thank Professor H.D.~Doebner for constructive discussions.
N.I.\ Stoilova was supported by project P6/02 of the Interuniversity Attraction Poles Programme (Belgian State -- 
Belgian Science Policy) and by the Humboldt Foundation.
R.\ Chakrabarti wishes to acknowledge Ghent University for a visitors grant.

\end{document}